\documentclass[aps,pre,twocolumn,10pt,
 notitlepage,
 floatfix,
 nofootinbib,
 superscriptaddress]{revtex4-2}
\usepackage{float}
\usepackage{subfigure}
\usepackage{amssymb}
\usepackage{amsfonts}
\usepackage{amsmath}
\usepackage{amsthm}
\usepackage{epsfig}
\usepackage{float}
\usepackage{multirow}
\usepackage[usenames,dvipsnames]{color}
\usepackage{xr}
\usepackage{enumitem}
\usepackage{braket}
\usepackage{graphics,graphicx,xcolor,subfigure}
\usepackage{comment}

 \usepackage[hidelinks,unicode=true]{hyperref}
 \hypersetup{colorlinks=true,
 	linkcolor=blue,
 	urlcolor=blue,
 	citecolor=blue,
 	pdfhighlight=/N
 }
 
\usepackage[normalem]{ulem}

\newcommand{\av}[1]{\langle {#1} \rangle}
\newcommand{\braketk}[1]{\left\langle {#1} \right\rangle_0}

\newcommand{\orcid}[1]{\hspace{0.2em}\href{https://orcid.org/#1}{\includegraphics[keepaspectratio,width=0.7em]{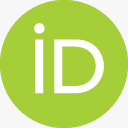}}}

\begin{document}



\title{
Strong localization blurs the criticality of time series for spreading phenomena on networks}


\author{Juliane T. Moraes\orcid{0000-0002-9199-8237}}
\affiliation{Departamento de F\'{\i}sica, Universidade Federal de Vi\c{c}osa, 36570-900 Vi\c{c}osa, Minas Gerais, Brazil}

\author{Silvio C. Ferreira\orcid{0000-0001-7159-2769}}
\affiliation{Departamento de F\'{\i}sica, Universidade Federal de Vi\c{c}osa, 36570-900 Vi\c{c}osa, Minas Gerais, Brazil}
\affiliation{National Institute of Science and Technology for Complex Systems, 22290-180, Rio de Janeiro, Brazil}

\begin{abstract}
We analyze critical time series of the order parameter generated with active to inactive phase transitions of spreading dynamics running on the top of heterogeneous networks. Different activation mechanisms that govern the dynamics near the critical point were investigated. The time series were analyzed using the visibility graph (VG) method where a disassortative degree correlation of the VG is a signature of criticality. In contrast, assortative correlation is associated with off-critical dynamics. The signature of criticality given by the VG is confirmed for collective activation phenomena, as in the case of homogeneous networks. Similarly, for a localized activation driven by a densely connected set of hubs, identified by a maximum $k$-core decomposition, critical times series were also successfully identified by the VG method. However, in the case of activation driven by sparsely distributed hubs, the time series criticality is blurred, being observable only for very large systems. In the case of strong structural localization induced by the presence of rare regions, an assortative VG degree correlation, typical of off-critical series,  is observed. We conclude that while macroscopic times series remain good proxies for the analysis of criticality for collective or maximum $k$-core activation, systems under spatial localization can postpone the signatures of or, in case of extreme localization, lead to false negatives for criticality of the time series.
\end{abstract}

\maketitle

\section{Introduction}

Criticality can be understood as a state of a large system of interacting agents lingering at the edge between order and disorder, in which nontrivial emerging phenomena are ubiquitous~\cite{Anderson1972,stanley1987introduction}. Formerly born in the realm of phase transitions in condensed matter physics at the thermodynamic equilibrium~\cite{stanley1987introduction}, critically has been expanded to a broad set of non-equilibrium~\cite{Tauber2014,Marro2005,Henkel2008} and natural systems~\cite{Sornet2006}. Several evidences indicate that criticality also plays a central role in biological systems~\cite{Munoz2017} as, for example, neuron firing dynamics on brain networks~\cite{Beggs2003, Chialvo2010, Kinouchi2006} and animal collective motion~\cite{Vicsek2012,Puy2024,Paiva2021}. Criticality presents universal features of scale invariance where, differently from noncritical states, an extensive range of space and time scales are relevant for emerging macroscopic properties~\cite{Stanley1999}. Scale invariance can manifest as critical exponents~\cite{Stanley1999,Henkel2008}, ageing effects~\cite{Tauber2014}, scale-free avalanches~\cite{Munoz2017}, and self-affine time series~\cite{Henkel2008,Peng1994,Goldberger2002,Barabasi1995}, the last being the central interest in the present work. 

The underlying structure where the agents interact to produce a critical behavior is commonly well represented by heterogeneous graphs~\cite{barabasibook} such as the structural or functional brain networks~\cite{Munoz2017,Bullmore2009,Eguiluz2005}, regulatory gene networks~\cite{Yeung2011}, epidemic spreading across technological infrastructures~\cite{Pastor-Satorras2015,Pastor-Satorras2001a}, human sexual contacts~\cite{Eames2002}, and geographical organization for human mobility~\cite{Danon2020,Costa2020}. Critical phenomena on networks enclose a wide class of statistical physics models and their corresponding analysis and methods~\cite{Dorogovtsev2008}. Dynamical processes on networks are characterized by nontrivial phenomena associated with activity localization induced by the network disorder~\cite{Munoz2010, Goltsev2012, Silva2021, Moretti2013, Arruda2017,Hebert2019} such as the presence of the hubs, cliques of densely connected nodes~\cite{Castellano2012,Kitsak2010}, and modular structures~\cite{Safari2017,Cota2018a}. Disorder and localization are known to play a central role in critical phenomena of low dimensional systems~\cite{Vojta2006}, being able to alter the nature of a critical phase transition introducing effects such as Griffiths phases~\cite{Vojta2014} or total smearing of critical point~\cite{Dickison2005}. In high dimensions, which is the case of the complex networks with small world property~\cite{barabasibook}, the disorder is not expected to play a relevant role for criticality in the thermodynamic limit~\cite{Munoz2010,Moretti2013} while relevant effects are present for finite sizes~\cite{Cota2016}. Therefore,  determining whether a system with the presence of heterogeneity is critical or not, is a nontrivial issue~\cite{Beggs2012}.

Several observables associated with dynamical processes are commonly expressed in the form of ordered time series~\cite{Shumway2011,Yanagawa2013,Yeung2011}. Recently, we~\cite{Moraes2023} applied the method of mapping time series of epidemic prevalence obtained from critical and noncritical spreading models on regular (homogeneous) lattices onto visibility graphs (VGs)~\cite{Lacasa2008,Lacasa2009}.  We reported that degree correlations are much more efficient in resolving critical and off-critical dynamics than the degree distribution~\cite{Moraes2023}. The method is very effective for high dimensions and, in particular, for infinite dimensional random regular networks. The hallmark of criticality is an asymptotic disassortative degree correlation where the higher the visibility of a time series point the lower the visibility of its connections. 

In the present work, we extend the analysis of critical prevalence (fraction of infected individuals) series for epidemic spreading models on highly heterogeneous networks, including scale-free and network with outliers for which different activation mechanisms are expected~\cite{Castellano2012,Kitsak2010,Sander2016}. Epidemic models on graphs are benchmarks of non-equilibrium critical phenomena through an absorbing state phase transition~\cite{Henkel2008,Pastor-Satorras2015,Dorogovtsev2008} and have been applied to investigate activation and inactivation dynamics in brain criticality~\cite{Kinouchi2006,Moretti2013}. We show that the VG method remains very effective for determining the critical behavior whether the activation mechanisms involve an extensive or subextensive component of the networks. However, the critical behavior is not evident if the spreading process is affected by strong localization as in networks with degree outliers or within a region of extended criticality in a Griffiths phase. We do not find false positives for criticality while false negatives happen in the presence of strong localization. So, the VG is an effective method to identify criticality in time series generated by very heterogeneous structures if it is not poisoned with strong localization.

The remainder of this paper is organized as follows. Section~\ref{sec:knn} presents an overview of the VG analysis of prevalence time series on a homogeneous substrate following Ref.~\cite{Moraes2023}. In section \ref{sec:nonlocal}, we analyze the critical series of simple spreading models where a collective activation is at work. In section \ref{sec:sis_pl}, we analyze the case of scale-free networks where activation involves subextensive components of the network. Strongly localized dynamics are tackled in section~\ref{sec:strong}. We present our concluding remarks in Sec.~\ref{sec:concl}.

\section{Methods}
\label{sec:knn}

\begin{figure}[!h]
    \centering
    \includegraphics[width=0.85\columnwidth]{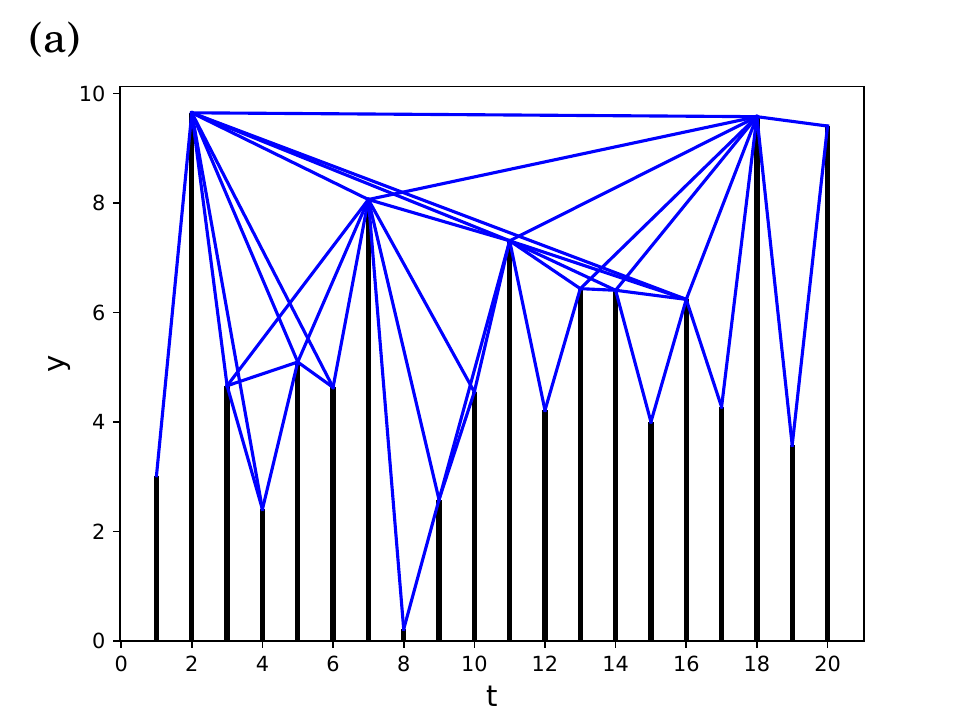}
    \includegraphics[width=0.85\columnwidth]{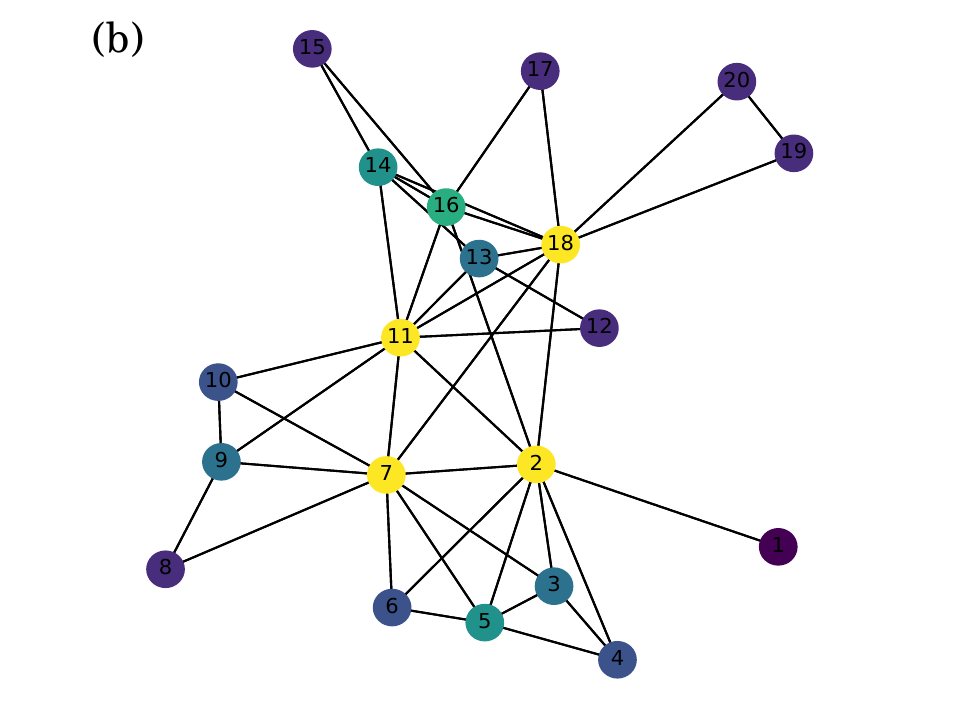}
    \caption{(a) Schematic representation of the visibility criterion of Eq.~\eqref{eq:visib} applied to a time series of 20 randomly generated values. (b) The visibility graph produced by the time series is shown in panel (a). }
    \label{fig:ex_vg}
\end{figure}

The visibility graph (VG) is a method for mapping a time series into an undirected graph~\cite{Zou2019}. In the present work, we use the natural VG proposed by Lacasa \textit{et al}.~\cite{Lacasa2008}, defined as follows. Considering an ordered time-series $\{(t_i,y_i)\}$, two points $(t_a,y_a)$ and $(t_b,y_b)$ are connected in the natural VG if all intermediate points $(t_c,y_c)$, where $t_a<t_c<t_b$, satisfies the visibility criterion given by
\begin{equation}
y_c < y_b + (y_a - y_b) \frac{(t_b - t_c)}{(t_b - t_a)}. 
\label{eq:visib}
\end{equation} 
Then, each point in the time series is associated with a node in the VG, which preserves the fractal structure of the time series~\cite{Lacasa2009,Lacasa2010}: random series are mapped onto graphs with exponential degree distributions while fractal series are mapped onto scale-free networks with degree exponents tied to the fractal dimension of the time series.  Figure~\ref{fig:ex_vg} illustrates a short time series and the corresponding natural VG.  Once the VG is generated, it is possible to use complex network tools to investigate the time series properties~\cite{barabasibook}, instead of standard analysis of fractal series~\cite{Shumway2011}. The VG method has been applied to the investigation of time series related to different problems such as Alzheimer's disease~\cite{Ahmadlou2010,Wang2016}, turbulence \cite{Liu2010,Iacobello2018} and classification of sleep stages~\cite{Zhu2012}. In the present work, we consider computer-generated time series using spreading processes where agents are represented by nodes and the interaction among them by the links of a network of size $N$. More specifically, a fluctuating order parameter related to the density of active nodes, the prevalence, is used to construct the time series to be analyzed via the VG method.

We emphasize that hereafter we deal with two types of networks with completely distinct meanings. One is the VG mapped from the time series while the other is the substrate where the spreading phenomena that generate the time series occur, the spreading network (SN). In the former, for instance, the degree represents the visibility of the time series point while in the latter the amount of contacts of a spreading agent. In order to avoid confusion, the VG properties are labeled with the subscript $_\text{vg}$ while the SN does not carry subscripts. For example, the degree of VG is represented by $k_\text{vg}$ while for SN  $k$, the size (number of components) of the VG  is $N_\text{vg}$ while for the SN is $N$,  and so on.

\begin{figure*}[t]
    \centering
\includegraphics[width=0.85\linewidth]{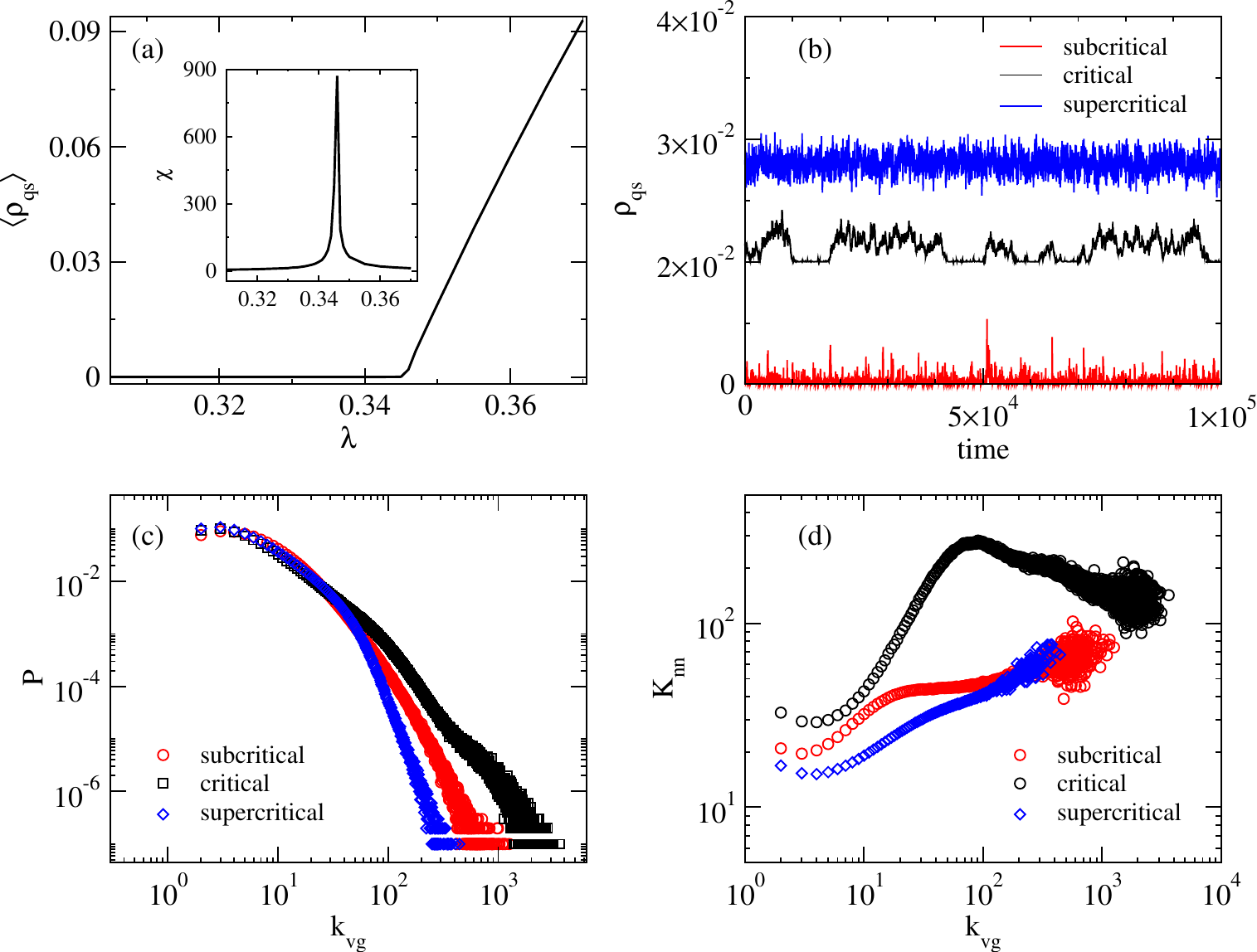}
    \caption{Analysis of the SIS model in a RRN of size $N=10^6$ and degree $k=4$ for all nodes. (a) Average quasistationary prevalence (the order parameter) as a function of the infection rate. The inset shows that the susceptibility as a function of the infection rate which presents a peak at the critical value $\lambda = \lambda_c$. (b) Quasistationary time series of epidemic prevalence in subcritical ($\lambda = 0.34$), critical ($\lambda = 0.346$), and supercritical ($\lambda = 0.352$) regimes. The time series were re-scaled and shifted to improve visualization. (c) Degree distributions and (d) average degrees of the nearest neighbors of the VG generated from the time series shown in (b). The curves in (c) and (d) are averages over 10 time series with $N_\text{vg}=10^6$ points separated in intervals $\delta t=1~\mu^{-1}$ for a single realization of an RRN. }
    \label{fig:example_rrn}
\end{figure*}

We will focus on epidemically motivated spreading models~\cite{Pastor-Satorras2015}, whereas the VG method applies to any dynamical process that produces time-ordered time series. We will consider two basic models with absorbing states: the susceptible-infected-susceptible (SIS)~\cite{Pastor-Satorras2001} and contact processes (CP)~\cite{Harris1974}. In both models, individuals can be in two states: susceptible (inactive) or infected (active). The active ones are spontaneously inactivated with rate $\mu$, fixed to $\mu=1$ by a choice of the time unit. In SIS dynamics an active node transmits the contagion to every inactive contact with rate $\lambda$ while in the CP  the infection rate is $\lambda/k$ per susceptible contact, where $k$ is the node degree in the SN. 

We performed stochastic simulations using the optimized Gillespie algorithm~\cite{Cota2017}. See Ref.~~\cite{Cota2017} for computer implementation of both SIS and CP models. The state where all individuals are inactive is absorbing for which the dynamics is trivially frozen. These states are handled by reactivating the most connected node in the heterogeneous case which allows to preserve the localization observed in standard quasistationary states~\cite{Sander2016}. The gold standard for quasistationary simulations~\cite{DeOliveira2005}, where the reactivation is done with a previously visited active configuration,  presents a drawback in which the reactivation leads to discontinuous time series and the VG  method does not work properly, especially in the critical region. In Ref.~\cite{Sander2016}, the temporal autocorrelation function and other relevant quantities of the standard quasistationary method were analyzed by removing the discontinuities. Results were found to be equivalent in the case of localized epidemic but algorithmically more complicated than hub reactivation. All the analyses are done after a relaxation time of $t_\text{rlx}=10^6~\mu^{-1}$.

The average activity prevalence in the quasistationary state is the order parameter. The transition point between absorbing and active phases can be numerically estimated through the dynamical susceptibility defined as $\chi=N[\av{\rho^2}-\av{\rho}^2]/\av{\rho}$~\cite{Ferreira2012}, computed in the quasistationary regime. As the system approaches the critical point, the susceptibility presents a maximum that diverges in the thermodynamical limit $N\rightarrow\infty$. Figure~\ref{fig:example_rrn}(a) illustrates typical prevalence and susceptibility curves for the SIS model evolving in a random regular network (RRN), in which all nodes have the same degree and the connections are randomly formed without multiple or self-connections~\cite{Ferreira2012}. Examples of time series generated for subcritical, critical, and supercritical regimes are presented in Fig.~\ref{fig:example_rrn}(b), where the fractal nature of the critical time series can be observed.  The VG degree distribution $P(k_\text{vg})$, defined as the probability of a VG node selected at random has degree $k_\text{vg}$~\cite{Lacasa2008}, are presented in Fig.~\ref{fig:example_rrn}(c). While a prominently heavier-tailed degree distribution is found for the critical case, no drastic differences between off-critical and critical regimes can be resolved in the curves.  

The average degree of nearest neighbors $K_\text{nn}$ as a function of the degree is a proxy to measure degree correlations in networks~\cite{Vazquez2002}. Assortative or disassortative behaviors happen when $K_\text{nn}$ is an increasing or decreasing function of the degree, indicating the propensity to connect with nodes of similar or dissimilar degrees, respectively~\cite{barabasibook}. We have shown that the critical behavior can be associated with a disassortative behavior of VG for large degrees $k_\text{vg}$ while an off-critical dynamics presents only the assortative regime~\cite{Moraes2023}. The degree correlations for VG generated by the times series shown in Fig.\ref{fig:example_rrn}(b) are presented in Fig.\ref{fig:example_rrn}(d) where one can clearly distinguish between the VG generated from the critical and off-critical time series through an asymptotic disassortative degree correlation for the former and pure assortativity for the latter.

The VG method is sensitive to the short time series lengths. A length analysis of critical series was presented in Ref.~\cite{Moraes2023} for delocalized dynamics where the degree correlations and distribution start to exhibit the asymptotic behavior for $N_\text{vg}>10^5$. In the present paper, a similar result is observed and hereafter the time series length is fixed to $N_\text{vg}=10^6$. The computational complexity of the natural VG method increases considerably if no optimization is adopted $O(N_\text{vg}^2)$. Our code, with intermediary optimizations, can compute natural VGs for $10^6$ points in one hour on a personal computer. If one considers $10^7$ points, the computer load increases to three days, still typically $O(N_\text{vg}^2)$.  Since criticality emerges for asymptotically large systems, we fix $N_\text{vg}=10^6$ points and perform a finite network size analysis which is sufficient for the current analysis.  If longer series are required, a computer complexity  $O(N_\text{vg}\ln N_\text{vg})$ can be achieved using a divide-and-conquer method of Ref.~\cite{Lan2015}.

\section{Spreading processes with collective activation mechanisms}\label{sec:nonlocal}

Spreading phenomena on heterogeneous networks can be triggered by activation mechanisms that involve different components of the network~\cite{Kitsak2010,Castellano2012,Goltsev2012,Ferreira2016,Cota2018}. When the activation is governed by a finite fraction of the network, the processes occur collectively~\cite{Ferreira2016,Cota2018} and behave as in a standard phase transition at a critical point. Here, one must differentiate between localized activation and activity: dynamical processes that depend explicitly on the contact structure of networks naturally present localization of activity due to the highly heterogeneous structure of connections, even if the activation is a collective phenomenon ~\cite{Silva2021}. Actually, the nodes with low activity, which comprise most of the network, represent the major contribution to the order parameter~\cite{Silva2021}.

\begin{figure}[!h]
    \includegraphics[width=0.9\columnwidth]{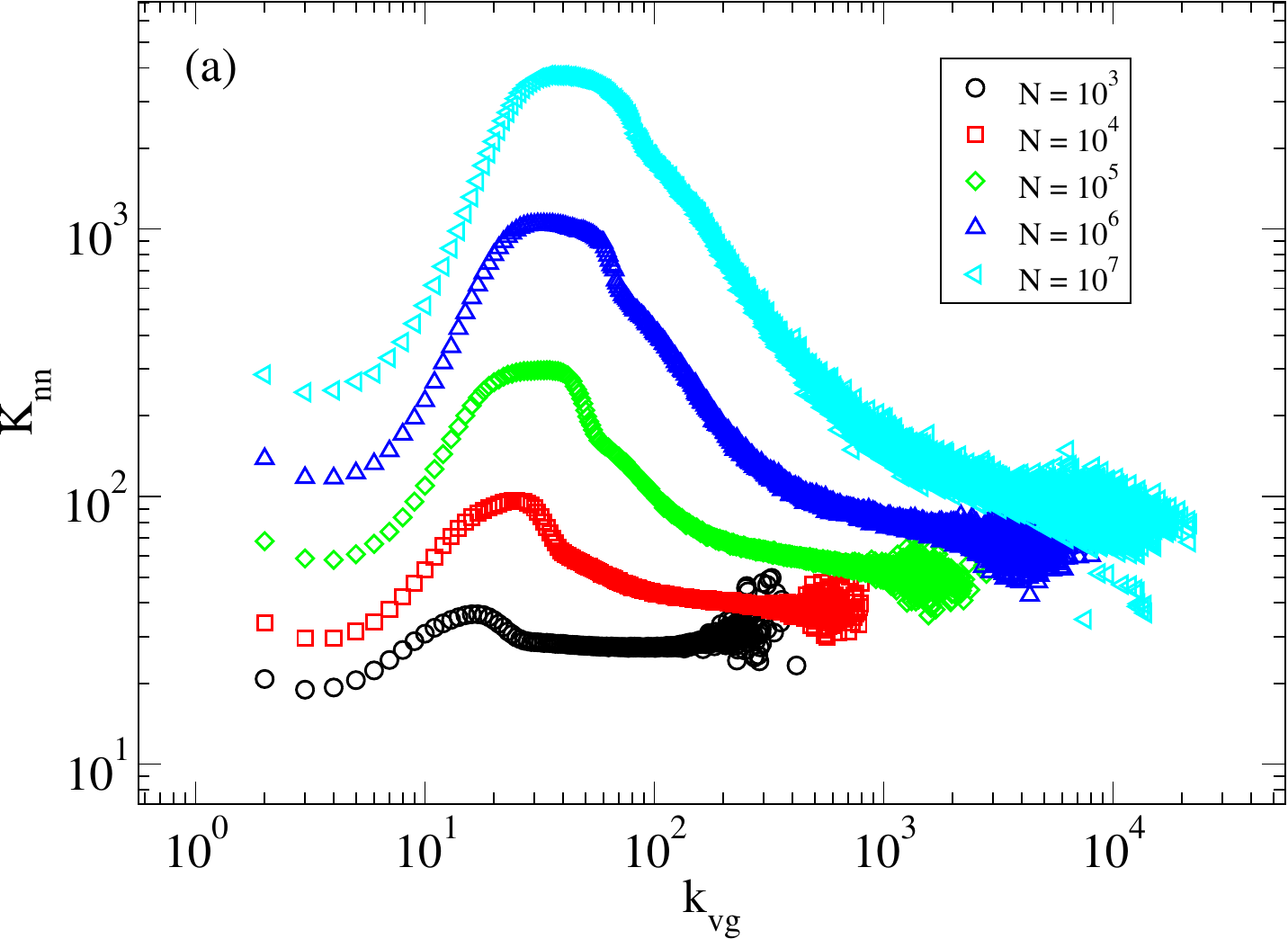}
    \includegraphics[width=0.9\columnwidth]{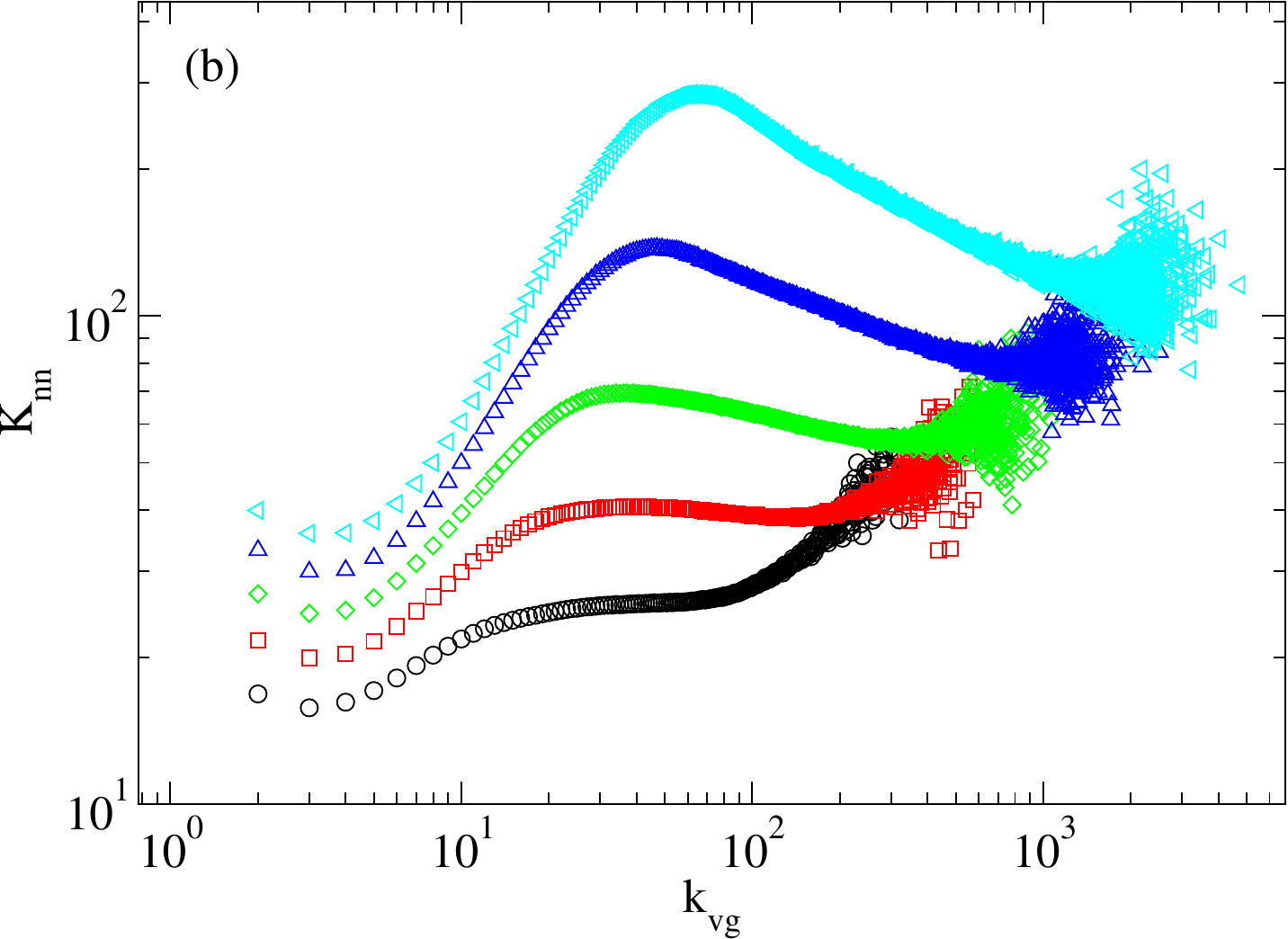}
    \caption{Average degree of the nearest neighbors of the VG generated from critical time series of the (a) SIS on annealed and (b)  CP model on quenched UCM networks of different sizes $N$, both with power-law degree distribution with exponent $\gamma = 2.75$, minimal degree $k_0=3$, and upper cutoff $k_\text{c}=2\sqrt{N}$. Averages over 10 time series of $N_\text{vg} = 10^6$ points were considered for a single SN realization}
    \label{fig:critical}
\end{figure}

Collective activation on heterogeneous networks is investigated considering SIS dynamics on annealed networks~\cite{Boguna2009} with power-law degree distribution, where the connections among nodes are probabilistic and rewired at every time step, implying that dynamical correlation between any pair of nodes is negligible in the thermodynamic limit. By construction, the activation is collective since all nodes interact with each other. The localized activation of SIS dynamics on quenched networks is considered in Section~\ref{sec:sis_pl}. We also investigate the contact process on quenched scale-free networks generated by the uncorrelated configuration model (UCM)~\cite{Catanzaro2005}, which undergoes a collective activation at a finite threshold~\cite{Ferreira2016}. Figure~\ref{fig:critical} shows the degree correlation analysis for the VG presenting an asymptotic disassortative behavior in $K_\text{nn}(k_\text{vg})$,  for both (a) SIS in an annealed and (b) CP in a quenched scale-free networks, respectively, becoming more evident as the network size increases. This signature is characteristic of critical time series in systems performing standard phase transitions~\cite{Moraes2023}. This result implies that the heterogeneity does not disturb the signature of criticality obtained from critical time series in the case of a collective activation.

It is worth noticing that the activation dynamics in the subcritical regime are sensitive to the quasistationary method used. For example, when a hub reactivation method is employed, a strong localization in a single node is artificially introduced irrespective of the model nature or network structure. If we consider reflecting boundary condition~\cite{Sander2016}, where the dynamics return to the state that it was immediately before falling into the absorbing state, the deep subcritical SIS dynamics present anomalous spiky time series in annealed networks. Indeed, when the dynamics are restarted in the hub of degree $k_\text{hub}$, the effective infection rate is $\lambda k_\text{hub} \gg 1$ in comparison with $\lambda \av{k} \ll 1$ for restarting in a typical node; the latter happens with higher probability than the former.  Nevertheless, this artifact in the subcritical region does not impair the analysis of the critical time series, which is the aim of this paper. Therefore, if there is no localization in the activation process we found that the critical time series provides VG with disassortative degree correlations, irrespective of heterogeneous SN.

\begin{figure*}[bht]
	\centering
	\includegraphics[width=1.9\columnwidth]{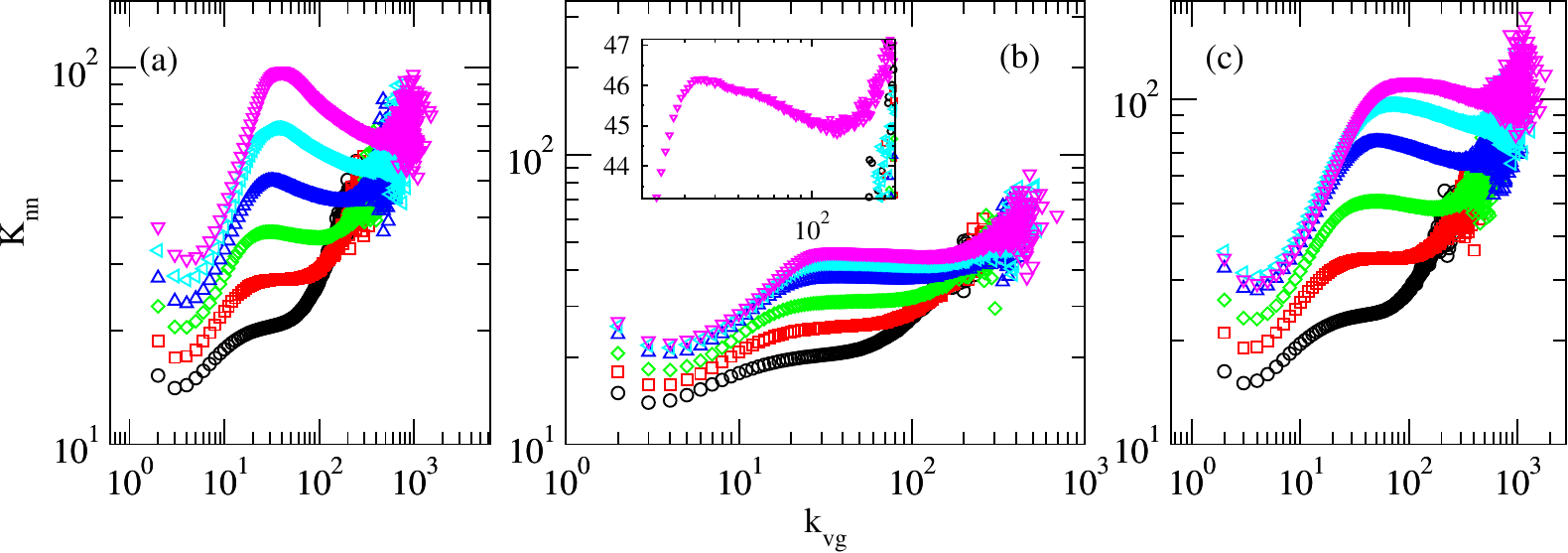}
	\caption{Average degree of the nearest neighbors for the VG generated from critical time series of prevalence for the  SIS model in UCM networks with (a) exponent $\gamma = 2.25$, (b) $\gamma = 2.75$, and (c)  $\gamma = 3.5$ and different sizes. The inset shows a zoom for the $N=10^8$ network with $\gamma=2.75$. For $\gamma<3$, a structural cutoff $k_\text{c}=2\sqrt{N}$ is used while a rigid cutoff $k_\text{c} = 3N^{1/\gamma}$ is considered for $\gamma=3.5$. The network sizes vary from $N = 10^3$ (black, bottom) to $N = 10^8$ (magenta, top). Averages over 5 samples of SN for each size and 10 time series with $N_\text{vg}=10^6$ points for SN were considered.}
	\label{fig:het_local}
\end{figure*}

\section{Spreading dynamics with localized activation mechanisms}\label{sec:sis_pl}

In contrast with the collective activation processes discussed in Sec.~\ref{sec:nonlocal}, other mechanisms, led by subextensive components of the network, are eligible~\cite{Kitsak2010,Castellano2012,Ferreira2016}. For example, the SIS dynamics on UCM networks with exponent $\gamma > 2.5$ are governed by long-range mutual infection of hubs separated a part~\cite{Castellano2012}. The hubs are sparsely distributed in the networks implying that the SIS dynamics produces loosely connected regions, composed of the hubs and their neighbors. There, the local epidemic activity holds for very long periods, exponentially large with hub degree,  permitting the infection propagation among hubs through paths of low degree nodes~\cite{Boguna2013}. 
For UCM networks with $\gamma<2.5$, the SIS activation is triggered in a densely connected set of hubs identified by the maximum $k$-core decomposition~\cite{Castellano2012}, which is a pruning process where peripheral nodes are sequentially removed up to the innermost core containing only nodes of degree $k\ge k_\text{core}$ is identified.  The interconnected hubs are activated due to their direct mutual interaction and they spread the epidemics out to the rest of the networks.

Figure~\ref{fig:het_local} presents the degree correlation analysis of VGs for critical dynamics for different regimes of SN degree exponent $\gamma$. For a maximum $k$-core activation, represented here by a UCM network with $\gamma=2.25$, the critical behavior is very similar to the collective case with a disassortative regime emerging asymptotically for large degrees. For mutual long-range activation of hubs, represented by a UCM network with $\gamma=2.75$, a neutral correlation pattern emerges for a broad range of sizes, while the asymptotic disassortative scaling is barely observed only in systems as large as $N=10^8$ nodes.  

\begin{figure}[hbt]
	\centering
	
	\includegraphics[width=0.95\columnwidth]{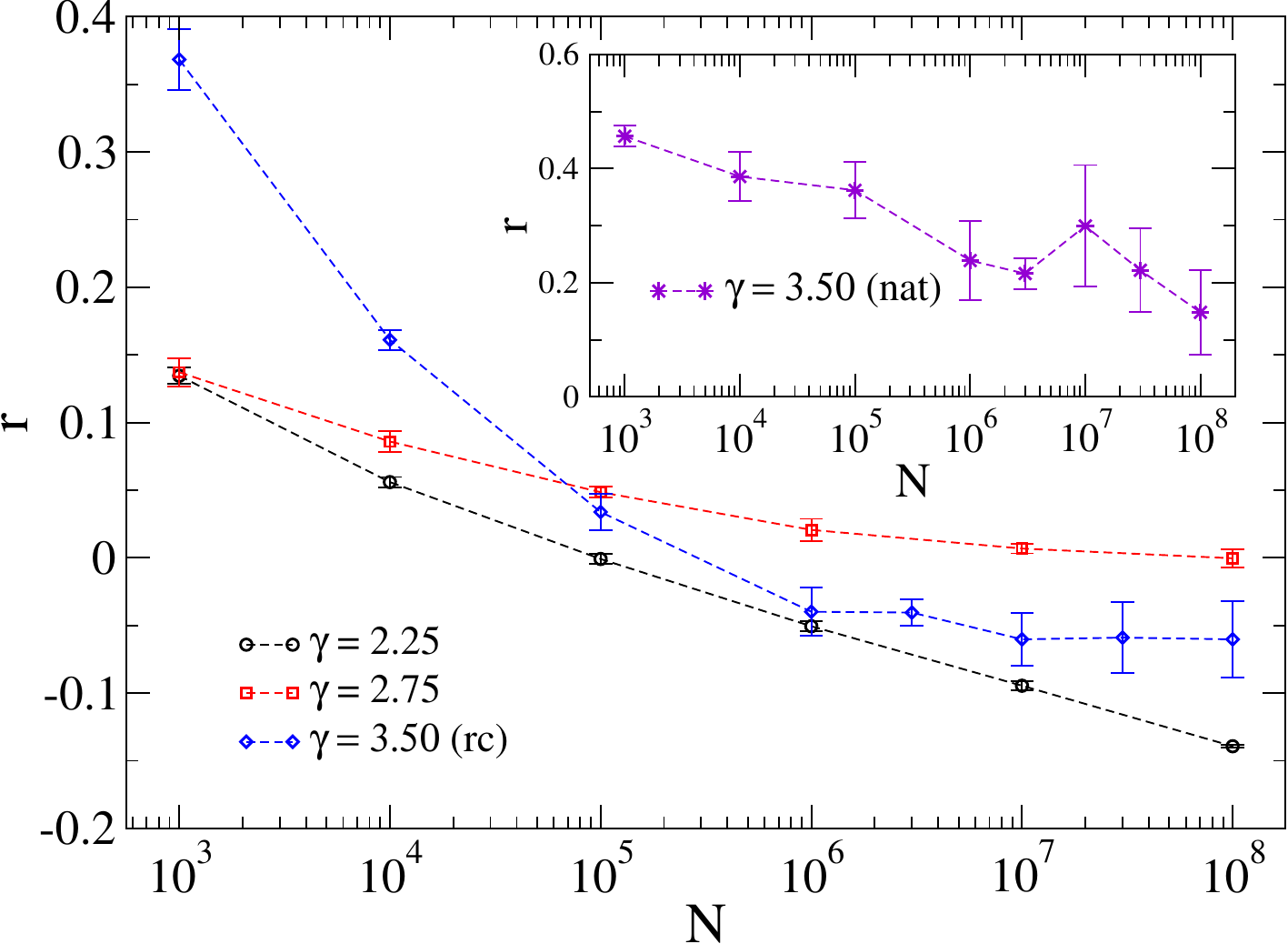}
	\caption{Pearson correlation coefficient of the VG as a function of the network size for critical time series obtained with SIS dynamics on UCM networks of different degree exponents. Structural ($k_\text{c}=2N^{1/2}$) and rigid cutoff ($k_\text{c} = 3N^{1/\gamma}$) were used for $\gamma<3$ and $\gamma>3$, respectively. Inset shows the VG corresponding to UCM network with $\gamma=3.5$ and natural cutoff. Averages over 5 samples of networks for each size and 10 time series with $N_\text{vg}=10^6$ points were considered.}
	\label{fig:pearson}
\end{figure}

For the UCM model with $\gamma>3$ one has outliers with degrees much larger than the remaining of the network due to the highly fluctuating natural cutoff $\av{k_{\max}}\sim N^{1/(\gamma-1)} \ll N^{1/2}$ that emerges in this regime~\cite{Boguna2009} and determining the epidemic threshold becomes cumbersome~\cite{Mata2015,Cota2016}.  So, we first analyzed the case $\gamma=3.5$ considering a rigid upper cutoff $k_\text{c}\sim N^{1/\gamma}$ for which maximum degree fluctuations are dropped substantially and an epidemic threshold can be unambiguously determined~\cite{Cota2016,Silva2019}. The case of natural cutoff is discussed in Sec.~\ref{sec:strong} in the context of strong localization phenomena. The degree correlation of the VGs obtained from critical time series is presented in Fig.~\ref{fig:het_local}(c) where a pattern similar to the case $\gamma=2.75$ is observed. The results presented in Figs.~\ref{fig:het_local}(b) and (c) suggest that activity's localization due to sparsely distributed hubs blurs the criticality signature 
in the form of a disassortative VG of the time series. The critical benchmark is barely observable only in the largest investigated sizes. We delve deeper into the localization effects in Sec.~\ref{sec:strong}.

While Fig.~\ref{fig:het_local} does not show clearly the criticality signatures for all values of $\gamma$, they are expected to emerge for sufficiently large systems. To analyze whether the asymptotic disassortative degree correlations of the VG emerge asymptotically, we computed a modified Pearson correlation coefficient ($r$)~\cite{Newman2002,Newman2010}. The $r$ values belong to the range $-1 \leq r \leq 1$, in which $r<0$ ($r>0$) indicates a disassortative (assortative) pattern in the network, while $r\approx 0$ represents a neutral correlation~\cite{Newman2002}. We define the partial Pearson coefficient determined  directly from  the degree distribution $P(q)$ and the average nearest neighbors degree $K_{nn}(q)$ as~\cite{barabasibook}:
\begin{equation}
r = \frac{1}{\sigma^2} \frac{\braketk{q^2}}{\braketk{q}}\left[\frac{\braketk{K_{nn}q^2}}{\braketk{K_{nn}q}} - \frac{\braketk{q^2}}{\braketk{q}} \right],
\end{equation}
in which 
\begin{equation}
	\sigma^2 = \frac{\braketk{q^3}}{\braketk{q}} - \frac{\braketk{q^2}^2}{\braketk{q}^2}.
\end{equation}
Here, $\braketk{\cdots}$ means partial average  constrained to $q>q_0$:
\begin{equation}
	\braketk{f} =\sum_{q=q_0}^{q_\text{max}} f(q) P(q)
\end{equation}
where $q_0$ is a lower bound chosen to exclude the lower degree nodes. The standard Pearson coefficient is recovered when $q=q_\text{min}$.

We analyze the finite-size scaling of the partial Pearson coefficient for VGs as a function of the UCM network size for curves presented in Fig.~\ref{fig:het_local}. Considering $q=k_\text{vg}$, $q_0 = 2\braket{k_\text{vg}}$ for scale-free UCM networks ($\gamma=2.25$ and $\gamma=2.75$), and $q_0 = 6\braket{k_\text{vg}}$  for the UCM networks with $\gamma=3.5$. The choice of this lower bound was based on the change of the pattern shown in Fig.\ref{fig:het_local} to exclude the initial assortative degree correlation regime common for all VGs, whether it is critical or not.  The partial Pearson coefficient analysis is presented in Fig.~\ref{fig:pearson}. On the one hand, for SIS on UCM networks with $\gamma=2.25$, a negative coefficient is obtained for UCM networks as small as $N \ge 10^5$ confirming criticality with a pattern similar to the collective activation; see Fig.~\ref{fig:critical}. On the other hand, in the case of mutual hub activation using UCM networks with $\gamma = 2.75$,  an extrapolation as  $r=r_\infty+aN^{-b}$ shows that the coefficient converges very slowly towards a value slightly negative $r_\infty \approx - 0.023$ for sizes $N>10^8$. For $\gamma=3.50$ with a rigid cutoff, a more disassortative coefficient $r_\infty \approx - 0.087$ is obtained while the case $\gamma=2.25$ a high disassortative VG with $r_\infty \approx - 0.37$ is found. The exponent $b$ used in the finite-size scaling is not universal. 

\begin{figure*}[hbt]
	\centering
	\includegraphics[width=0.85\linewidth]{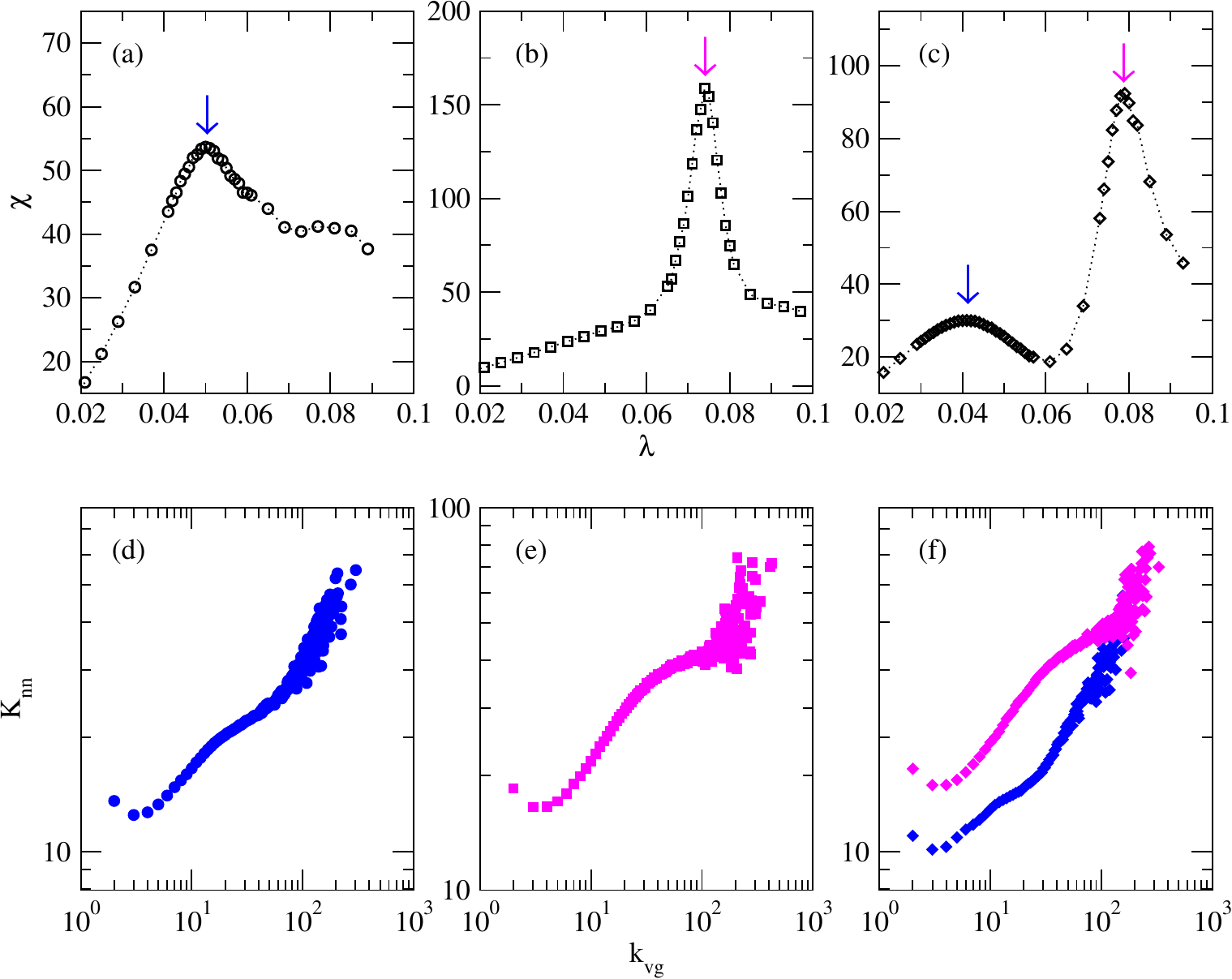}
	\caption{Top panels: susceptibility as a function of infection rate for SIS model in 3 different samples of UCM networks with degree exponent $\gamma = 3.5$, a natural cutoff and size $N = 10^7$. Bottom panels: average degree of the nearest neighbors for the VGs generated from the time series computed at the infection rates given by the susceptibility peaks indicated by the arrows in the top panels. A total of 10 time series with $N_\text{vg}=10^6$ points were considered for single SN realizations. }
	\label{fig:ucm3-5}
\end{figure*}

\section{Spreading dynamics with strong localization}\label{sec:strong}

The weak signature of criticality of VG obtained for the SIS on SNs with $\gamma>2.5$, suggests that localization can alter the signatures of criticality in time series. So, we now investigated dynamics with strong localization.

\subsection{SIS on UCM networks with $\gamma=3.50$}

UCM networks with $\gamma>3$ and natural degree cutoff possess outliers (few nodes of a degree much larger than the others in the network) sparsely distributed that lead to independently active domains in the subcritical phase that can be detected as multiple peaks in the susceptibility curves~\cite{Mata2015}. Fig.~\ref{fig:ucm3-5} shows the analysis of times series obtained for 3 different samples of UCM networks with $N=10^7$ nodes, degree exponent $\gamma = 3.5$, and a natural cutoff. The top panels show susceptibility curves and the peaks indicated by the arrows. In Fig.~\ref{fig:ucm3-5}(c), the susceptibility presents two well-resolved peaks; the first one is related to the activation of an outlier, and the second one is related to the global activation of the network. The bottom panels of Fig.~\ref{fig:ucm3-5} show the degree correlations analysis of the VG generated with the prevalence time series at infection rates corresponding to the peaks of susceptibility. Only assortative regimes emerge in these cases, which were previously associated with noncritical dynamics, even when the system is in a global activation threshold. Notice, however, that even in the absence of outliers shown in Fig.~\ref{fig:ucm3-5}(b), the disassortative asymptotic behavior of critical series is not observed. The partial Pearson coefficient, pictured in the inset of Fig.\ref{fig:pearson}, has large fluctuations being not clear whether it follows a decay towards a negative Pearson coefficient (disassortative behavior) or not. So, the localized activity produced by outliers smears the criticality signatures, at least for networks as large as $N=10^8$.

\subsection{SIS on a RRN with an outlier}

\begin{figure}[hbt]
	\centering
	\includegraphics[width=\linewidth]{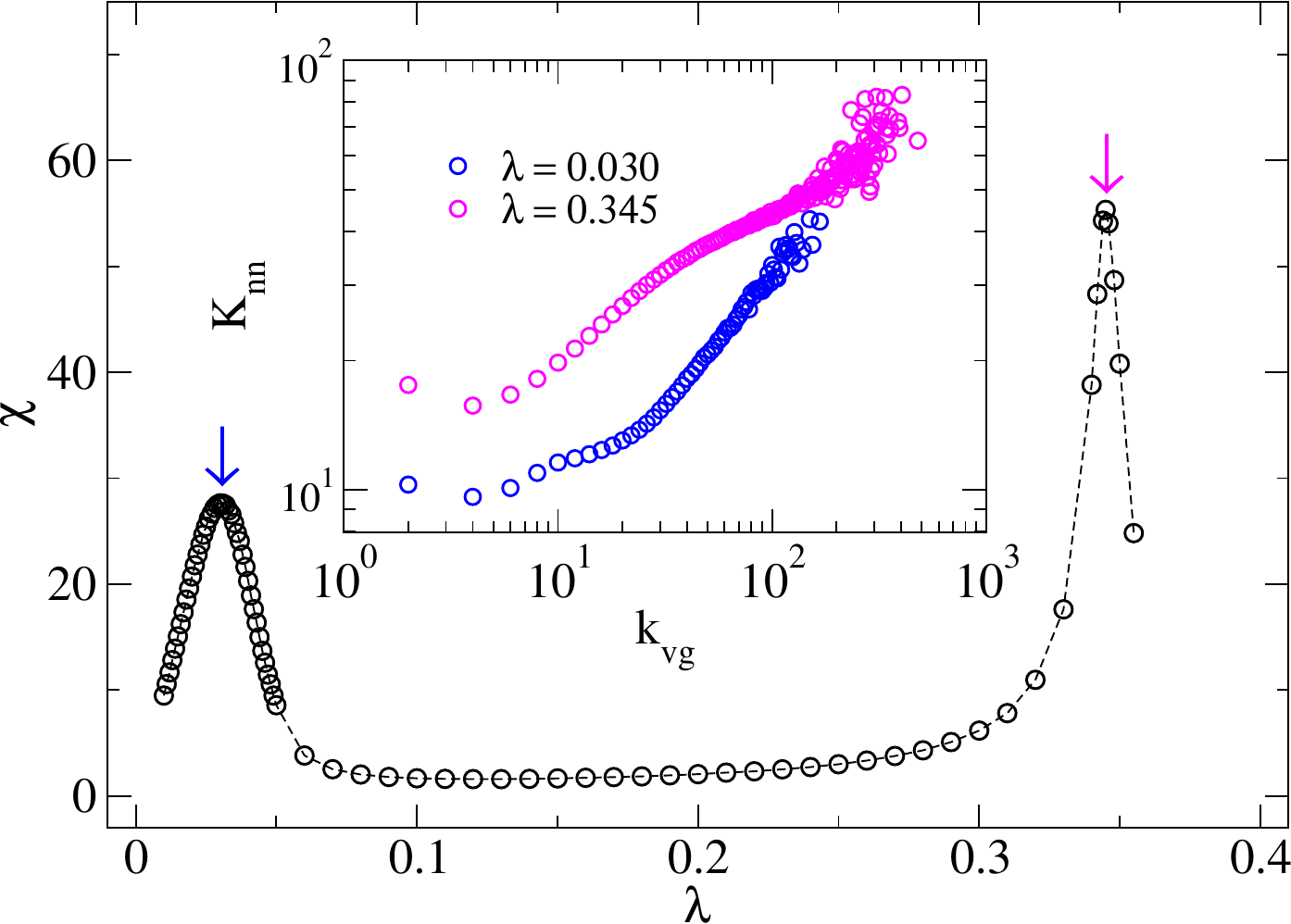}
	\caption{Main plot: susceptibility as a function of the infection rate for the SIS dynamics on a RRN of size $N=10^7$  where all nodes have degree $k=4$ except a hub with degree $k_\text{hub}=\sqrt{N}$.  Inset: average degree of the nearest neighbors for the VG generated from the prevalence time series computed with the infection rates corresponding to peaks of susceptibility shown in the main panel. A total of 10 time series with $N_\text{vg}=10^6$ points were considered for single RRN realizations.}
	\label{fig:rrn_hub}
\end{figure}

To shed light on this discussion, we consider a simpler graph for spreading dynamics composed of an RRN with a hub~\cite{rsferreira2016} where all nodes have the same degree $k$ except one, the hub, that has a degree $k \ll k_\text{hub}\ll N$. To mimic the UCM model, we chose $k_\text{hub}=\sqrt{N}$.  In these simplified networks, the actual endemic phase with finite epidemic prevalence emerges collectively when the pure RRN is activated and a clear separation between local and global activation can be controlled, as seen in Fig.~\ref{fig:rrn_hub}. While, as expected, the local activation of the hub in subcritical phase presents the assortative correlation patterns characteristic of off-critical dynamics, the critical pattern changes drastically as compared with the SIS dynamics in a pure RRN [ Fig.\ref{fig:example_rrn}(d)] as shown in the inset of Fig.~\ref{fig:rrn_hub}. Therefore, the criticality signature given by the VG degree correlations is blurred by the localization provided by the hub. This means that even if the time series is due to a critical state, the degree correlation of the VG will not present the disassortative behavior characteristic of clean systems if the dynamics in the underlying SN present very strong localization.

\subsection{Contact processes on diluted networks }

The disorder can induce strong localization that may replace a critical point by an extended region of criticality forming Griffiths phases~\cite{Vojta2006}, which can appear on networks if the disorder is sufficiently strong to produce independently activated regions~\cite{Munoz2010}. The extended region implies nonuniversal power-law decays of the order parameter within a finite range of the control parameter. We investigated the diluted contact process where the infection rate of each node $i$ follows a bimodal distribution $\lambda_i=0$ with probability $p$ and $\lambda_i=\lambda$ with complementary probability $1-p$~\cite{Munoz2010}. The null infection rate of a node is equivalent to a dilution.

\begin{figure}[hbt]
    \centering
    
    \includegraphics[width=\columnwidth]{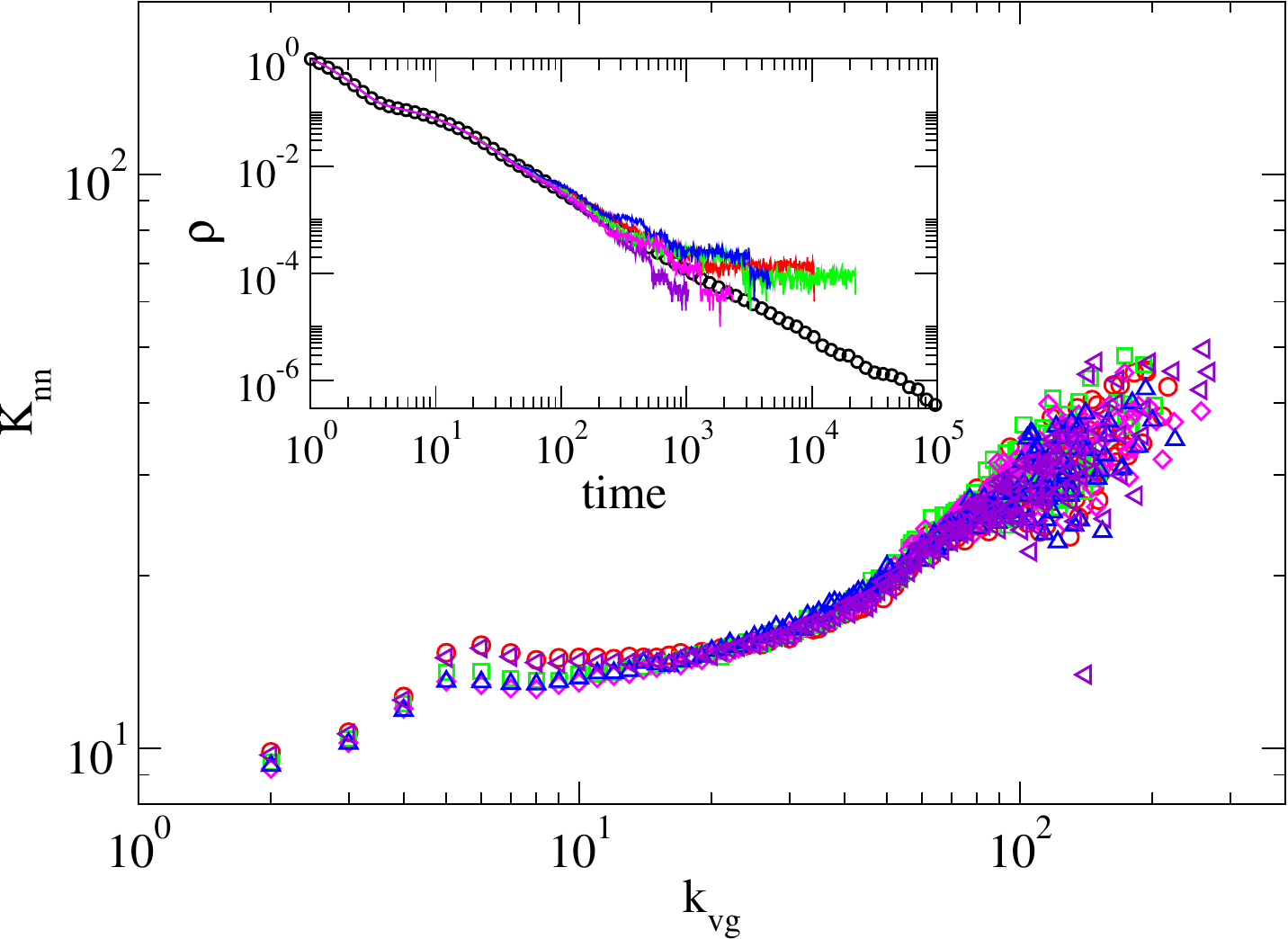}
    \caption{Main plot: average degree of the nearest neighbors for the VG generated from the quasistationary time series of prevalence for the 5 disorder realizations shown in the inset. Inset: prevalence decay over time for a diluted contact process in an Erdos-Renyi network with size $N = 10^5$ and average degree $\braket{k} = 3$ using an infection rate of $\lambda = 20$ and dilution $p = 0.9$. The symbols are an average over $10^3$ dilution realizations, which present the power-law decay, while curves depicted by lines of different colors show the decay for 5 randomly chosen dilution realizations.  }
    \label{fig:qcp_decay}
\end{figure}

We consider Erd\"os-Renyi (ER) networks with $\braket{k} = 3$, $N = 10^5$ nodes as SN. A dilution parameter $p=0.9$, which produces very clear extended regions presenting power-law decay, is used; see Fig.1 of Ref.~\cite{Munoz2010}. The inset of  Fig.\ref{fig:qcp_decay} shows the power-law decay of the fraction of active nodes for $\lambda=20$, considering a fully active initial conditions and an average of over $10^3$ dilution realizations. Also, $5$ curves illustrating the decay for single disorder realizations are shown. The degree correlations of the VG generated for the quasistationary time series with these $5$ disorder realizations are shown in the main plot,  where one sees that all are strongly assortative. While the critical CP dynamics provided VG with disassortative degree correlations for all investigated networks, irrespectively of the underlying SN, the disorder in the order parameter changes significantly the assortativity pattern of VG in analogy with the strong localization caused by outliers in Figs~\ref{fig:ucm3-5} and \ref{fig:rrn_hub}. In the case of the diluted contact processes,  we have that the dilution fraction $p$ is above the percolation threshold of ER networks and the rare regions are formed by a group of nodes surrounded by diluted nodes.

\section{Conclusion}\label{sec:concl}

The determination of whether a complex system is critical or not demands a careful analysis since heterogeneity inherent to them can alter the nature of the transition~\cite{Munoz2017}. Recently, we have shown that the mapping of time series of the order parameter into visibility graphs can successfully differentiate critical and off-critical dynamics on spreading phenomena on regular lattices, the more effective for higher dimensions~\cite{Moraes2023}.  The central signature of criticality or lack of it is assortativity (off-critical) or disassortative (critical) asymptotic behaviors manifested in degree correlation analysis of the visibility graphs. In this work, we applied the method to analyze different dynamical processes near to the active to inactive phase transitions in distinct network structures, presenting high heterogeneity and different activation mechanisms.

Two types of activation mechanisms, involving subextensive components of the underlying network, were addressed considering the SIS dynamics. For a maximum $k$-core activation, the VG analysis of the critical time series can identify criticality, and it is very similar to the case of collective activation which involves an extensive part of the system. For mutual activation of sparsely distributed hubs the scenario changes. When localization is reduced by avoiding the presence of outliers, the criticality signature is postponed for very large systems. However, when degree outliers are allowed leading to a strong localization around the hubs and their nearest neighbors, the criticality signature disappears even in the case of a single hub immersed in a sea of low-degree nodes. Finally, we investigate another type of strong localization where the dynamics present Griffiths phases with power-law decay of the order parameter within an extended region of criticality. We report a similar behavior of the SIS dynamics in the presence of outliers.

In a nutshell, while the VG method introduced in Ref.~\cite{Moraes2023} remains effective in determining the time series criticality in the presence of structural heterogeneity if localization is not extreme, as it does the delocalized cases, we observed that strong localization can blur the signatures of the criticality of the time series, even when the system is at an actual critical point. The last assertion becomes evident for a random regular network (RRN) with a single hub shown in Fig.~\ref{fig:rrn_hub}, where the global transition happens at the same threshold of the pure RRN, while a local activation occurs in the subcritical phase due to the hub. It is important to stress that the VG method did not provide a false positive for criticality, that is, every time the asymptotic disassortative behavior was observed, the system was indeed critical. However, false negatives occur whenever the critical systems are under strong localization effects.

\appendix

\begin{acknowledgments}
S.C.F. acknowledges the financial support by the \textit{Conselho Nacional de Desenvolvimento Científico e Tecnológico} (CNPq)-Brazil (Grant no. 310984/2023-8) and \textit{Fundação de Amparo à Pesquisa do Estado de Minas Gerais} (FAPEMIG)-Brazil (Grant No. APQ-01973-24).  This study was financed in part by the \textit{Coordenação de Aperfeiçoamento de Pessoal de Nível Superior} (CAPES), Brazil, Finance Code 001.
\end{acknowledgments}


%

\end{document}